\documentstyle[aps,prl]{revtex}

\begin{document}
\twocolumn[
\title{ Wigner crystallization and metal-insulator transition of two-dimensional holes in GaAs at 
B=0 }
\author{ Jongsoo Yoon, C.C. Li, D. Shahar$^{1}$, D.C. Tsui, and M. Shayegan }
\address{Department of Electrical Engineering, Princeton University, Princeton, NJ08544}
\address{$^{1}$Department of Condensed Matter Physics, Weizmann Institute, Rehobot 76100, Israel }
\date{July 14, 1998}

\begin{center}
\parbox{15cm}{\small
We report the transport properties of a low disorder two-dimensional hole system (2DHS) in the 
GaAs/AlGaAs heterostructure, which has an unprecedentedly high peak mobility of $7\times 10^5$cm$^2$/Vs, with hole density of $4.8\times 10^9$cm$^{-2}$$<$p$<$3.72$\times$10$^{10}$cm$^{-2}$ in the temperature range of  50mK$<$T$<$1.3K. 
From their T, p, and electric field dependences, we find that the metal-insulator transition in zero 
magnetic field in this exceptionally clean 2DHS occurs at $r_s=35.1\pm0.9$, which is in 
good agreement with the critical $r_s$ for Wigner crystallization ${r_s}^c=37\pm 5$, 
predicted by Tanatar and Ceperley for an ideally clean 2D system.

}
\end{center}

\maketitle
]

It is well known that the ground state of a two-dimensional (2D) electron or hole system realized 
in semiconductors at B=0 is expected to be a Wigner crystal (WC) in the low density limit 
when carrier-carrier interaction becomes dominant over the kinetic energy of the carriers. 
Tanatar and Ceperley \cite{1} from their Monte Carlo simulation for ideally clean systems find 
that the critical density, expressed in terms of the dimensionless parameter $r_s$, is 
$r_s=37\pm 5$. We note that $r_s$ is the Coulomb energy to Fermi energy ratio, given by 
$r_s=(\mbox {p}\pi)^{-1/2}m^{*}e^2/4\pi \hbar^2 \epsilon \epsilon _o$, where $m^{*}$ is the 
effective mass, p the charge carrier density, and $\epsilon$ the dielectric constant. Disorder, which 
is always present in any experimental system, also plays important roles. In the clean limit, it is 
expected to pin the WC, rendering the system into an insulator. In the highly disordered dirty 
limit, the Coulomb interaction is negligible, or can be treated as a small perturbation; the system 
is better described by single particle localization. In between, when both interaction and 
localization are equally important, neither the electronic processes nor the nature of the phase is 
known and there is no clear physical picture for the system at the present time. Also, it is 
believed that if the disorder is not too strong, it can enhance Wigner crystallization. Chui and 
Tanatar \cite{2} find from their Monte Carlo studies that the disorder they put in their model can 
stabilize the WC phase at $r_s$ as low as 7.5.

On the experimental side, a number of publications \cite{3,4,5,6,7,8,9,10}, since the discovery by 
Kravchenko et al. \cite{3} of a 2D metal-insulator transition (MIT) in high mobility Si metal-
oxide-semiconductor field-effect transistors (Si MOSFET's) at $r_s \approx 10$, have reported 
transport studies in the large $r_s$ regime of 2D electron or hole systems in several different 
semiconductor heterostructures. These studies generally do not address the properties in the 
insulating phase. They focus mostly on the nature of the metallic phase and the scaling behavior 
near the MIT. Their results agree with the conclusion of Kravchenko et al. \cite{3} that there is a 
MIT in high mobility samples and the critical density for the transition varies from $r_s\approx 
5$ in $\mbox {Si}_{0.88}\mbox {Ge}_{0.12}$ to $r_s\approx 28$ in p-GaAs. The origin and nature of the 
insulating phase, however, can be inferred from earlier Si-MOSFET experiments, where the low 
mobility samples show a cross-over from weak localization to strong localization as the carrier 
density is decreased. In high mobility samples, which show MIT's at $r_s\approx 10$, Pudalov et 
al. \cite{11} first studied the transport characteristics in the insulating phase and attributed them 
to those of a pinned WC. However, Shashkin et al. \cite{12} and Mason et al. \cite{13} 
subsequently find that single particle localization gives a better description and the transport is 
via hopping in the presence of a Coulomb gap. It thus appears that in the regime where disorder 
and localization must be treated on equal footing, transport and I-V characteristics do not allow 
us to identify whether the insulating phase is a WC. A more desirable approach to the problem 
has to come from clear identification of a 2D system in the clean limit, where the physics is 
expected to be simpler and theoretical predictions more reliable.

In this letter, we report transport measurements on a 2D hole system (2DHS) in 
GaAs/Al$_{x}$Ga$_{1-x}$As, which has an unprecedentedly high peak mobility, $\mu _p=7\times 10^5 
\mbox {cm}^2$/Vs. This mobility, approximately twice time higher than previously studied 
2DHS, allowed us to reach p as low as $4.8\times 10^9 \mbox {cm}^{-2}$. From a systematic 
study of density, temperature and bias dependences of the hole transport, we find a MIT at 
$r_s=35.1\pm 0.9$, which is in good agreement with $r_s=37\pm 5$ predicted for 
the WC transition by Tanatar and Ceperley. We also examine the dependence of $r_s$ at the MIT, 
from all 2D electron and hole systems that have been reported to show MIT's, on the disorder in 
the sample and show that our sample is indeed in the clean limit. In the rest of this paper, we 
present this result and also describe in detail related new transport characteristics from our 
2DHS.

The 2DHS is created in a modulation doped GaAs/Al$_{x}$ Ga$_{1-x}$As heterostructure grown on 
the (311){\em A} surface of GaAs substrate by molecular-beam epitaxy. A backgate, used to change 
the density, is situated 250$\mu$m away from the 2DHS, and does not screen the carrier-carrier 
interaction. Since the holes are produced by modulation doping, the background neutralizing 
charges are immobile. In contrast, the 2D carriers in the insulated gate field-effect transistors 
used in other experiments \cite{3,4,5,6,7} are capacitively induced by applying a voltage to a metallic 
gate  located close to the 2D  systems.   In  such  transistors,   the background  neutralizing 
charges are mobile and the carrier-carrier interaction is screened.
The sample was a rectangle of $4.2\times 0.4\mbox {mm}^2$ cut along the $[\overline{2} 33]$ direction, 
after the wafer was thinned to 250$\mu$m. Resistivity ($\rho$) was measured by a four probe 
method. To avoid sample heating, current was limited to below 1nA, and voltage was measured 
across two contacts separated by 0.7mm located near the center. An ac technique using a lock-in 
at a frequency $<$10Hz was used whenever possible. At low densities where capacitive coupling 
became significant, a dc method was employed. The sample was cooled in the mixing chamber 
of a dilution refrigerator. Three samples cut from the same wafer were studied, and all showed 
MIT's with the same characteristics. However, the data reported in this Letter were collected over 
three cool-downs on the same sample. The sample was warmed up to room temperature between 
cool-downs.



Fig. 1 shows the T dependence of $\rho$, measured in the zero current limit, in the range 
50mK$<$T$<$1.3K, for $4.8\times 10^9\le p\le 3.72\times 10^{10}\mbox{ cm}^{-2}$ or 
$16.0\le r_s\le 44.4$. For p$>$2$\times$ 10$^{10}$cm$^{-2}$ (the two bottom traces), the sign of 
d$\rho$/dT, which has been usually used to distinguish metallic phase from the insulating phase, is 
positive in the entire T range, showing metallic behavior. For p$\le$7.2$\times$10$^{9}$cm$^{-2}$ (the 
four top traces), d$\rho$/dT is always negative exhibiting the characteristic of an insulator. In the 
intermediate range of p, however, there is a peak in $\rho$ resulting from d$\rho$/dT$<$0 at high T 
and d$\rho$/dT$>$0 at low T. This peak starts to appear at p=1.98$\times$10$^{10}$cm$^{-2}$; it 
becomes more pronounced and shifts to a lower temperature as p decreases. The maximum is at 
T$\approx$0.7K at p=$1.98\times 10^{10}\mbox{cm}^{-2}$, and $\approx$0.18K at p=$9.0\times 10^9  
\mbox{cm}^{-2}$. The presence of such a peak makes it ambiguous to tell whether the 2DHS is in the 
metallic phase or insulating phase using the sign of d$\rho$/dT as the criterion. For example, at 
p=$9.0\times 10^9\mbox{cm}^{-2}$, if the measurements are made only down to 0.3K, one would 
conclude from the fact d$\rho$/dT is negative that the 2DHS is in the insulating phase, while the 
data below 0.18K indicate that the 2DHS is in fact in the metallic phase.


However, this ambiguity can be avoided if we define the phase as metallic when the I-V is 
superlinear (d$^{2}$I/dV$^{2}$$<$0) and insulating when it is sublinear (d$^{2}$I/dV$^{2}$$>$0). Fig. 2 is the 
differential conductance (dI/dV) plotted against V at p=$9.0\times 10^9 \mbox{cm}^{-2}$, taken at 
several temperatures below and above the $\rho$ maximum, as marked by the arrows in the inset. 
In sharp contrast to d$\rho$/dT, which shows a metallic behavior below and an insulating behavior 
above T=0.18K, the dI/dV vs. V taken at all four temperatures show superlinear I-V's, i.e., 
dI/dV decreases with increasing V. Such a superlinear I-V is observed down to p=$8.1\times 
10^9 \mbox{cm}^{-2}$. When p is reduced below $7.3\times 10^9 \mbox{cm}^{-2}$, where $\rho$ shows 
monotonic increase with decreasing T, the I-V becomes sublinear, indicative of an insulating 
phase.  It should also be pointed out that identifying the metallic phase and the insulating phase 
based on the sign of d$^{2}$I/dV$^{2}$ gives exactly the same result as that obtained using the sign of 
d$\rho$/dT at our lowest T.

Using the sign of d$^{2}$I/dV$^{2}$, we locate the MIT at p=$7.7\pm 0.4\times 10^9 \mbox{cm}^{-2}$, or 
equivalently at $r_s=35.1\pm 0.9$. At the MIT, $\rho$ is approximately 35k$\Omega$ at 
T$\rightarrow$0, and $E_F$ is calculated to be 0.58K. The critical $r_s$ for the MIT in our sample is approximately four times that in high mobility Si MOSFET's and 25\% larger than that in 
previously studied highest mobility 2DHS in GaAs/AlGaAs, of which $\mu_{p}\approx 4\times 10^5 
\mbox {cm}^2$/Vs is approximately half of that in our 2DHS. It is clear from this result that the $r_s$, at 
which the 2D MIT is observed, depends strongly on the disorder in the 2D system. In the inset of 
Fig. 1, we plot ${r_s}^c$, the 2D MIT critical $r_s$, from all samples reported in the literature as a 
function of the scattering rate of the 2D carriers, defined by $1/\tau =e/(m^{*}\mu _p)$, which 
gives a simple measure of the disorder and also takes into account of the differences in effective 
mass, $m^{*}$, of the charge carriers in different material systems. It is clear that for 1/$\tau$ 
$>$1$\times$ 10$^{11}$ sec$^{-1}$, where all the high mobility Si MOSFET data lie, ${r_s}^c$ is 
approximately constant. The data from the 2D systems in GaAs/AlGaAs heterostructures, which 
are known to have low disorder and show fractional quantum Hall effect characteristics in high 
magnetic fields, are all in the 1/$\tau$$<$1$\times$10$^{11}$sec$^{-1}$ regime. In this low scattering 
rate region, ${r_s}^c$ shows a steep increase with decreasing 1/$\tau$  and extrapolates to 
${r_s}^c$ $\approx$40 as 1/$\tau$$\rightarrow$0. Our data, shown as the filled circle in the figure is in the clean limit and the observed ${r_s}^c=35.1\pm 0.9$ is in good agreement with the 
$r_s=37\pm 5$ for the WC transition predicted by Tanatar and Ceperley.

In the clean limit, a WC is expected to be pinned, giving rise to a depinning threshold voltage, 
which is characterized by a sudden increase in current as the applied voltage exceeds the 
threshold. In Fig. 3, the dV/dI taken at $r_s$=44.4 is plotted against V. At 50mK, the dV/dI is 
nearly constant until V reaches about 1.5mV, beyond which it drops rapidly by an order of 
magnitude at V$\approx$3mV. This is suggestive of the depinning threshold expected for a pinned 
WC. In the WC conduction model proposed by Chui \cite{14}, where the WC is weakly pinned 
by the potential of remote dopants, transport is expected to be mediated by the creation of 
dislocation pairs through quantum tunneling. The energy needed to create a dislocation pair leads 
to a conduction threshold voltage. The observed threshold voltage of 1.5mV, corresponding to a 
field of 2.1V/m, is in reasonable agreement with Chui's model, according to which, the 
depinning threshold field estimated for our sample is $E_d=0.09{n_i}a^2e^2/\epsilon 
d^3=2.5$V/m. Here, $n_i=8\times 10^{11} \mbox {cm}^{-2}$ is the remote doping impurity density, 
$a=(\pi \mbox {p})^{-1/2}$, and the doping impurity setback distance d=1200 $\AA$. However, 
there are caveats. The pinning gap deduced from $E_d$, $\Delta _p=\hbar (eE_d/m^{*}a)^{1/2}\approx$2.7K, is not consistent with the observation that no well defined depinning 
threshold feature is seen when the thermal energy is still much smaller than the gap. In Fig. 3, the 
data show no clear threshold at T as low as 145mK. Moreover, as shown in the inset of Fig. 3, 
the I-V taken at T as low as 50mK is linear within the experimental resolution at voltages below 
1.5mV. This ohmic conduction below the depinning threshold voltage cannot be explained by 
possible backgate leakage current, which is much less than 1pA.

Finally, we turn to the overall symmetry in the nonlinear transport properties of our low disorder 
2DHS as its density is varied across the MIT. Shahar et al. \cite{15} recently reported a new 
symmetry across the quantum Hall liquid to Hall insulator transition. They find that the I-V 
curves in the quantum Hall liquid phase can be mapped into the I-V curves in the Hall insulator 
phase by simply interchanging I with V. This I-V inversion symmetry can be viewed as 
reflecting the charge-flux duality symmetry in the composite boson description of the fractional 
quantum Hall effect. More recently, Kravchenko et al. \cite{4} and Simonian et al. \cite{16} 
reported a similar reflection symmetry across the 2D MIT, though its origin is still unclear at the 
present time. We have made a detailed comparison of the I-V characteristics in the metallic and 
insulating phases in our sample, and find that they are qualitatively different. Instead of 
comparing I-V with V-I in the two phases, we compare their derivatives which are more 
sensitive to changes in the curvature. In Fig. 4(a) and 4(b), dI/dV in the insulating phase and 
dV/dI in the metallic phase are shown against I and V, respectively. In the metallic phase, as V 
increases, dV/dI initially increases, but stops increasing at V$\approx$ 0.15mV. Above this 
characteristic voltage, dV/dI stays constant to I as high as 100nA, the highest current used in our 
measurements. If we assume a reflection symmetry with respect to a straight line in I-V 
corresponding to $\rho$=35k$\Omega$/sqr. at the MIT, the characteristic voltage of 0.15mV in 
the metallic phase implies a characteristic current of 2.5nA in the insulating phase, above which 
dI/dV should remain constant. It is clear from Fig. 4(a) that such a characteristic current is absent 
in all traces of the dI/dV vs. I in the insulating phase. We conclude that the I-V reflection 
symmetry across the 2D MIT is not observed in our data in the ranges of I and V shown in Fig. 
4.

To summarize, we presented transport properties of a 2DHS that undergoes a MIT at 
${r_s}^c=35.1\pm 0.9$, which is unambiguously determined using the sign of d$^{2}$I/dV$^{2}$ rather 
than d$\rho$/dT. The dependence of ${r_s}^c$ on disorder, from all 2D electron and hole systems 
that have been reported to show MIT's, indicates that our 2DHS is in the clean limit, and the 
value of ${r_s}^c$ in our sample is in good agreement with $r_s=37\pm 5$ expected for the WC 
transition by Tanatar and Ceperley. In the insulating phase, a conduction threshold that is 
quantitatively consistent with the WC conduction model of Chui is found, but the pinning gap 
estimated from the observed threshold field is too large compared with our observations. We also 
find that the I-V reflection symmetry across the MIT is absent in our low disorder 2DHS.

We thank R. Bhatt, S. Chakravarty, M. Hilke, Y. Hanein, and S. Papadakis for fruitful 
discussions. This work was supported by the NSF.


\vspace{.5cm}
Fig. 1.  T dependence of $\rho$, from the top curve, at p=0.48, 0.55, 0.64, 0.72, 0.90, 1.02, 1.27, 1.98, 
2.72, and 3.72 in units of 10$^{10}$ cm$^{-2}$. The inset shows the dependence of ${r_s}^c$ on 1/$\tau$, from various 2D systems; this work (filled circle), 2DHS in GaAs/AlGaAs (open circles. \cite{7,8,9}), 2DES 
AlAs/GaAlAs (filled triangle, \cite{6}), 2DHS in Si$_{0.88}$Ge$_{0.12}$/Si (filled square, \cite{10}), and 2DES in Si-MOSFET (open triangles, \cite{3,4,5}). $m^{*}$/$m_e$=0.37, 0.46, 0.18, and 0.19 for p-GaAs \cite{17}, n-AlAs \cite{18}, p-Si$_{0.88}$Ge$_{0.12}$ \cite{19}, and n-Si-MOSFET.
\vspace{.5cm}

Fig. 2. dI/dV in the metallic phase near the MIT (p=9.0$\times$$10^9$ cm$^{-2}$) at four different temperatures marked by the arrows in the inset. Qualitatively same nonlinear I-V characteristics are seen at all four temperatures. In contrast, d$\rho$/dT shows a metallic behavior below and an insulating behavior 
above T=0.18K.
\vspace{.5cm}
 
Fig. 3. dV/dI at p=4.8$\times$10$^9$ cm$^{-2}$ is plotted against V at T=50mK (open circles), 145mK (filled 
circles), and 284mK (triangles). The inset shows the I-V curve at T=50mK. 
\vspace{.5cm}

Fig. 4. (a) dI/dV vs. I in the insulating phase and (b) dV/dI vs. V in the metallic phase at 50mK. 
The numbers shown next to the symbols are the densities in units of 10$^{10}$ cm$^{-2}$. The 
corresponding rs values are in parenthesis.


\begin{thebibliography}{99}

\bibitem{1} B. Tanatar and D.M. Ceperley, Phys. Rev. B {\bf 39}, 5005 (1989)
\bibitem{2} S.T. Chui and B. Tanatar, Phys. Rev. Lett. {\bf 74}, 458 (1995)
\bibitem{3} S.V. Kravchenko, G.V. Kravchenko, J.E. Furneaux, V.M. Pudalov, and M. D'Iorio, Phys. 
Rev. B {\bf 50}, 8039 (1994)
\bibitem{4} S.V. Kravchenko, D. Simonian, and M.P Sarachik, Phys. Rev. Lett. {\bf 77}, 4938 (1996)
\bibitem{5} D. Popovic, A.B. Fowler, and S. Wash burn, Phys. Rev. Lett. {\bf 79}, 1543 (1997); S.V. 
Kravchenko, Whitney E. Mason, G.E. Bowker, J.E. Furneaux, V.M. Pudalov, and M. D'Iorio, 
Phys. Rev. B {\bf 51}, 7038 (1995); V.M. Pudalov, cond-matt/970076; D. Simonian, S.V. 
Kravchenko, M.P. Sarachik, and V.M. Pudalov, Phys. Rev. Lett. {\bf 79}, 2304 (1997); V.M. Pudalov, 
G. Brunthaler, A. Prinz, and G. Bauer, cond-matt/9707054
\bibitem{6} S.J. Papadakis and M. Shayegan, Phys. Rev. B {\bf 57}, R15068 (1998)
\bibitem{7} Y. Hanein, U. Meirav, D. Shahar, C.C. Li, D.C. Tsui, and H. Shtrikman, Phys. Rev. Lett. {\bf 80}, 
1288 (1998)
\bibitem{8} M.Y. Simmons, A.R. Hamilton, M. Pepper, E.H. Linfield, R.D. Rose, D.A. Ritchie, A.K. 
Savchenko, and T.G. Griffiths, Phys. Rev. Lett. {\bf 80}, 1292 (1998)
\bibitem{9} M.Y. Simmons, A.R. Hamilton, T.G. Griffiths, A.K. Savchenko, M. Pepper, and D.A. 
Ritchie, cond-matt/9710111
\bibitem{10} P.T. Coleridge, R.L. Williams, Y. Feng, and P. Zawadzki, cond-matt/9708118
\bibitem{11} V.M. Pudalov, M. D'Iorio, S.V. Kravchenko, and J.W. Campbell, Phys. Rev. Lett. {\bf 70}, 1866 
(1993)
\bibitem{12} A.A. Shashkin, V.T. Dolgopolov, and G.V. Kravchenko, Phys. Rev. B {\bf 49}, 14486 (1994)
\bibitem{13} W. Mason, S.V. Kravchenko, and J.E. Furneaux, Surf. Sci. {\bf 361/362}, 953 (1996)
\bibitem{14} S.T. Chui, Phys. Lett. A {\bf 180}, 149 (1993); S.T. Chui, J. Phys. Cond. Matt. {\bf 5}, L405 (1993)
\bibitem{15} D. Shahar, D.C. Tsui, M. Shayegan, E. Shimshoni, and S.L. Sondhi, Science {\bf 274}, 589 
(1996)
\bibitem{16} D. Simonian, S.V. Kravchenko, and M.P. Sarachik, Phys. Rev. B {\bf 55}, R13421 (1997)
\bibitem{17} K. Hirakawa, Y. Zao, M.B. Santos, M. Shayegan, and D.C. Tsui, Phys. Rev. B {\bf 47}, 4076 
(1993)
\bibitem{18} T.S. Lay, J.J. Heremans, Y.W. Suen, M.B. Santos, K. Hirakawa, M. Shayegan, and Zrenner, 
Appl. Phys. Lett. {\bf 62}, 3120 (1993)
\bibitem{19} S.-H. Song, D.C. Tsui, and F.F. Fang, Solid State Comm. {\bf 96}, 61 (1995) 


\end{thebibliography}
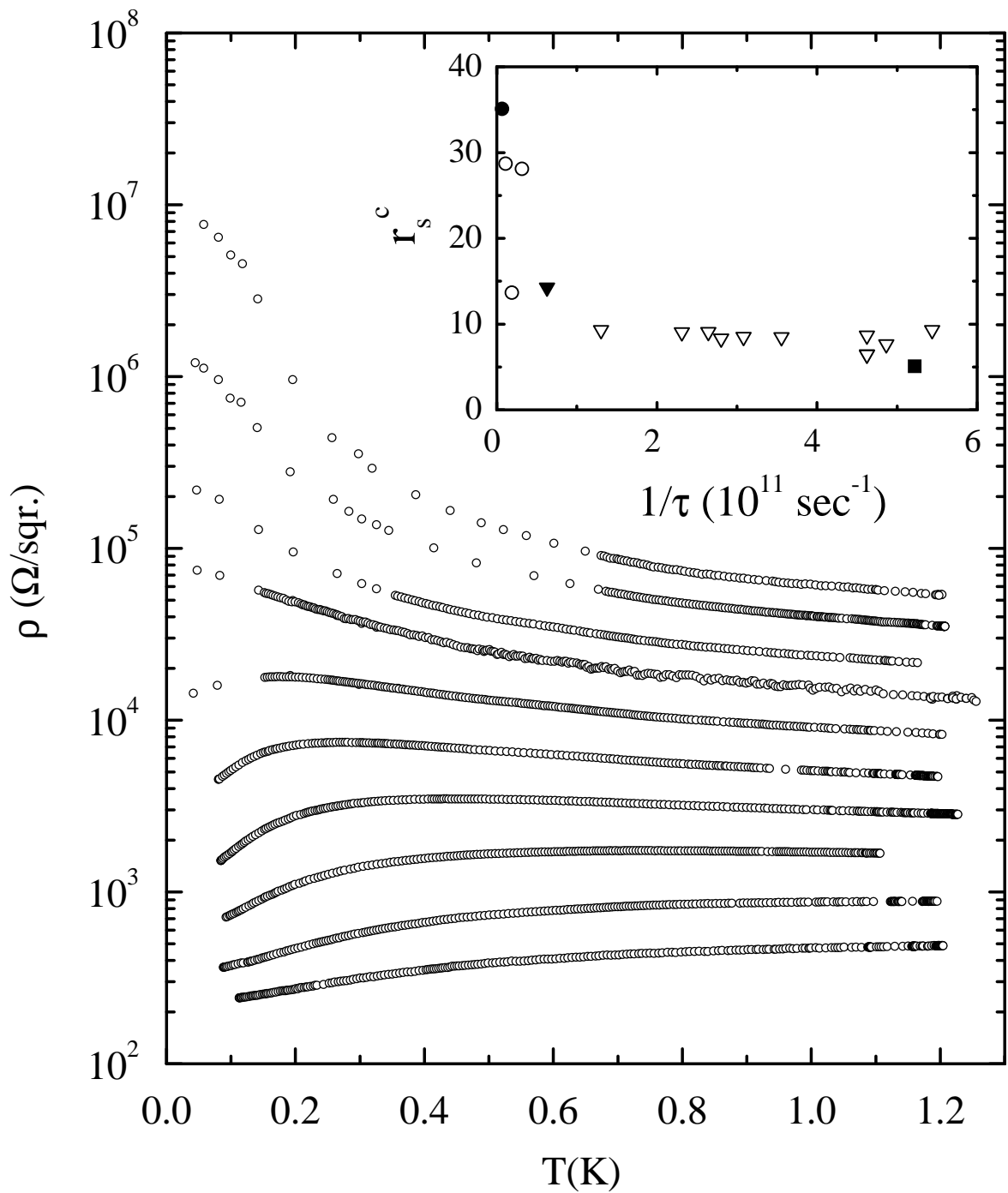
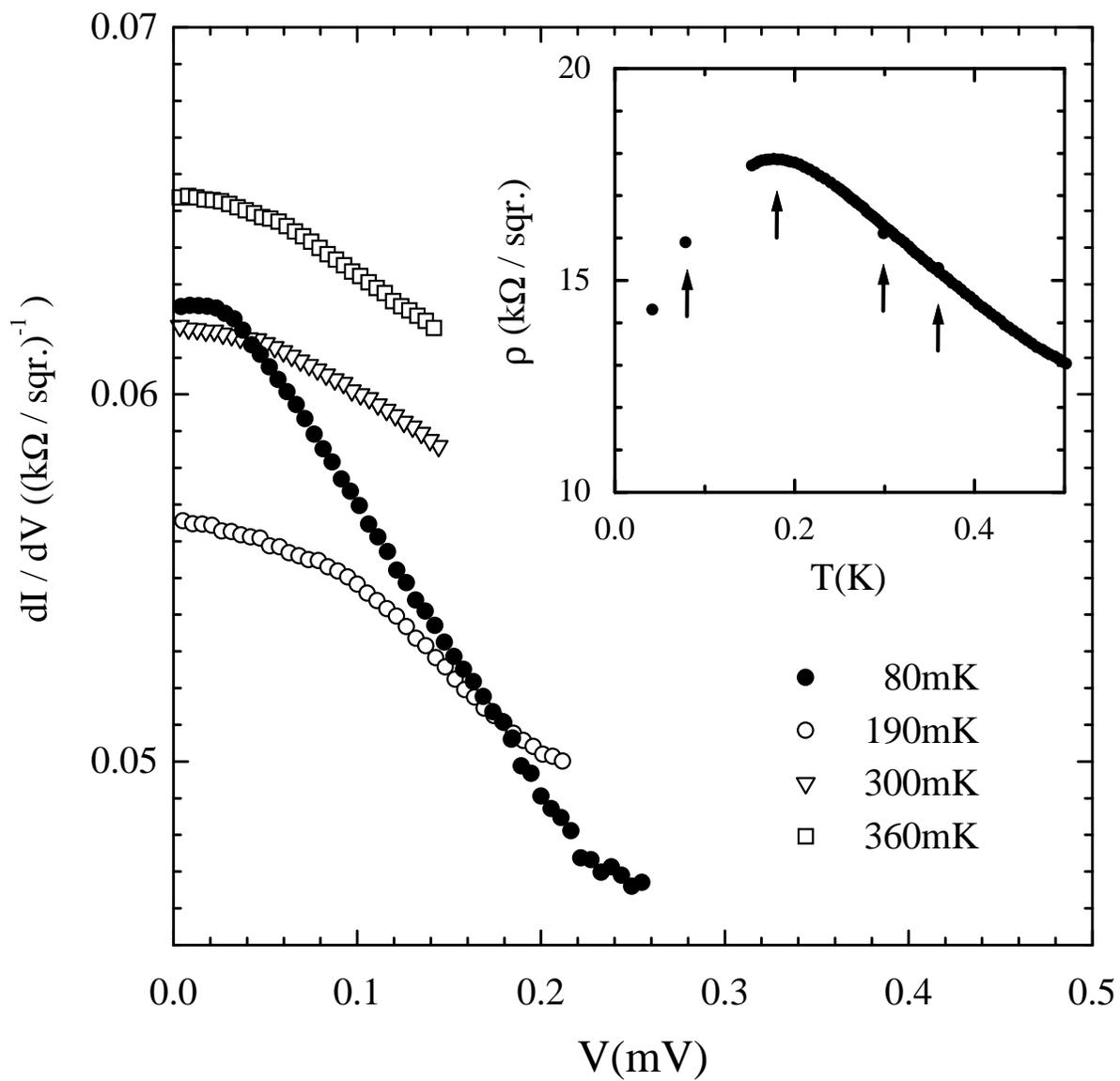
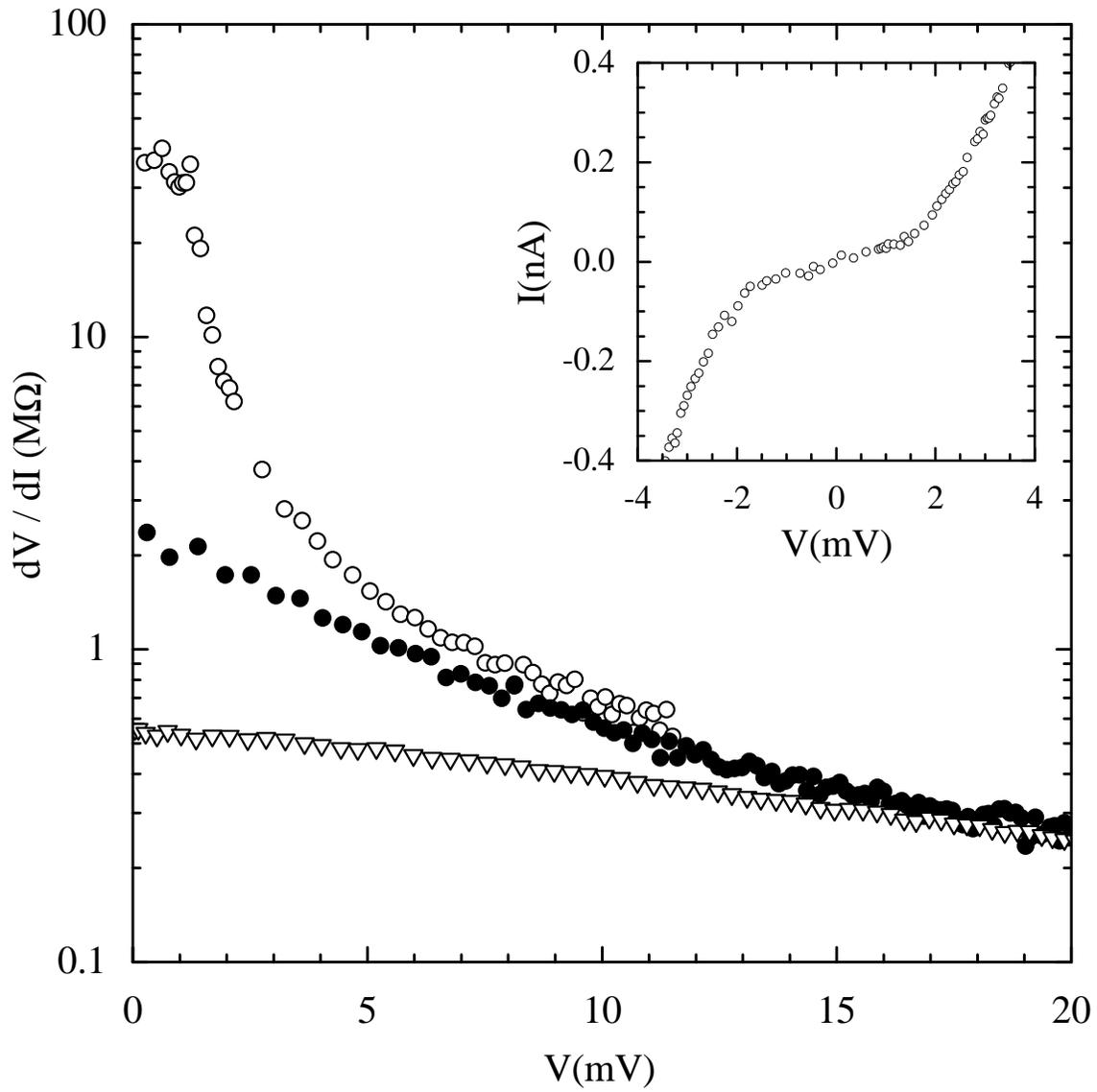
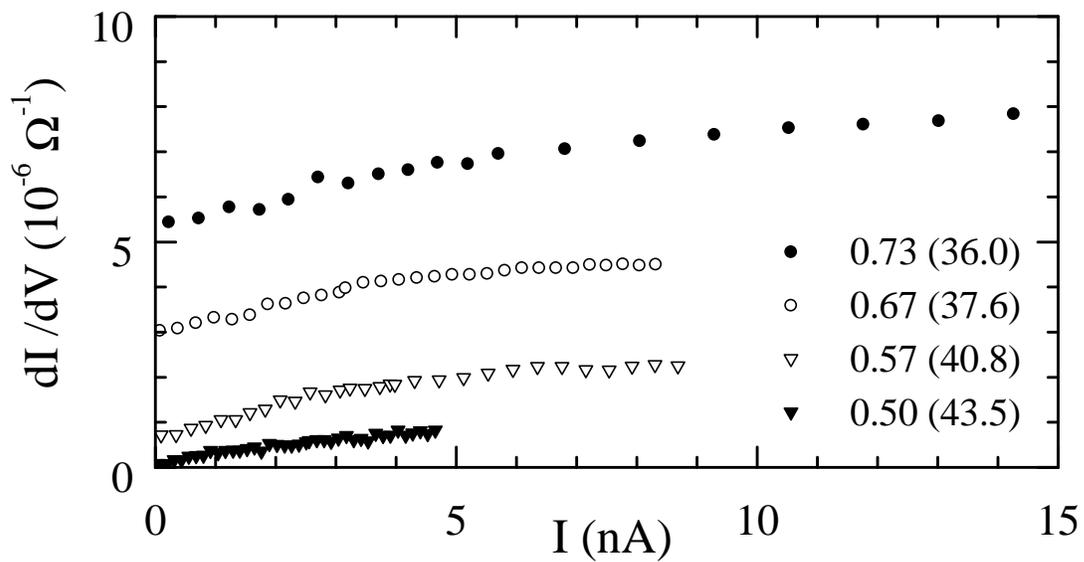
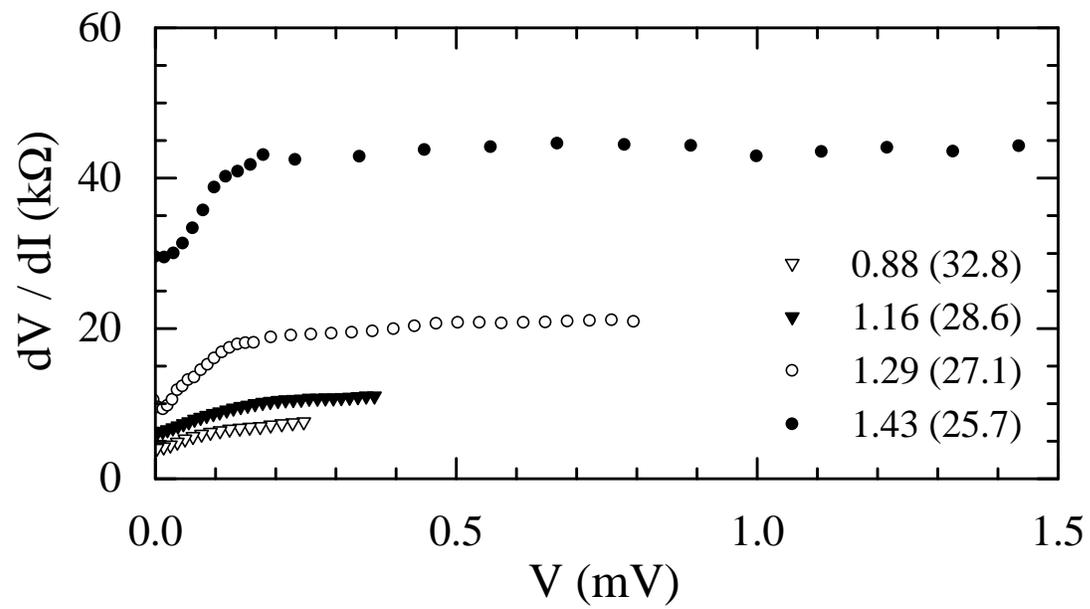
\end{document}